\documentclass[conference]{IEEEtran}
\IEEEoverridecommandlockouts

\usepackage{cite}
\usepackage{amsmath,amssymb,amsfonts}
\usepackage[T1]{fontenc}
\usepackage{algorithmic}
\usepackage{graphicx}
\usepackage{textcomp}
\usepackage{xcolor}
\usepackage{booktabs}
\usepackage{multirow}
\usepackage{url} 
\usepackage{verbatim}
\usepackage{soul}

\def\BibTeX{{\rm B\kern-.05em{\sc i\kern-.025em b}\kern-.08em
    T\kern-.1667em\lower.7ex\hbox{E}\kern-.125emX}}

\begin{document}

\title{Pedestrian Trajectory Prediction \\ through LoRA Federated Training \\
\thanks{This work is partly supported by the FAR 2025 Grant.}
}
\author{\IEEEauthorblockN{Simone Travagliati and Maria Luisa Merani}
\IEEEauthorblockA{
\textit{Department of Engineering “Enzo Ferrari''} \\
\textit{University of Modena and Reggio Emilia}, Modena, Italy \\
simone.travagliati@unimore.it, marialuisa.merani@unimore.it}
}
\maketitle

\begin{abstract}
In Intelligent Transport Systems, protecting vulnerable road users, specifically pedestrians, is a paramount challenge. The present work contributes to this field, 
putting forth a distributed architecture in which vehicles connected to the cellular infrastructure collect pedestrians' trajectories and fine-tune a foundation model for pedestrian trajectory forecasting via federated training.
A server positioned at the edge of the cellular network is responsible for aggregating the local model's parameters from the vehicles and updating the global model's parameters. To reduce radio resource utilization, Low-Rank Adaptation is proposed; to decrease both the number and the size of updates, an increase in the number of training rounds performed by the vehicles per communication round is put forth. A novel figure of merit is then introduced to quantify the use of radio resources dedicated to Vehicle-to-Infrastructure communications as required by the federated solution.

The proposed scheme delivers excellent prediction accuracy while significantly reducing the communication cost compared to alternative federated training methods. This shows that combining fine-tuning, Low-Rank Adaptation, and a reduction of the communication periodicity between the vehicles and the edge server is key to enabling a distributed approach to trajectory prediction in vehicular environments.  
\end{abstract}


\section{Introduction}

In the road safety domain, predicting pedestrian trajectories is one of the most complex tasks to solve, demanding extremely small errors that cannot be guaranteed without the adoption of AI-based models fed by large datasets.  
Traditionally, the task of trajectory prediction heavily relied on knowledge-based approaches, which, while providing valuable insights, often struggle to capture the complexity of human gait.
On the contrary, data-driven approaches have the ability to learn complex patterns from data that are difficult to be described by physical models and have recently gained prominence in trajectory prediction tasks, especially for pedestrians \cite{pedestrian_trajectory_forecasting_survey}.

Collecting pedestrians' data via vehicles 
is the natural choice, with Federated Learning (FL) representing the most attractive option for privacy-aware model training.
However, state-of-the-art complex neural networks would require the wireless transmission of large amounts of data, often hundreds of megabytes or even gigabytes, during each FL training round, prohibiting the model construction and its update in real time. 
Federated Fine-Tuning (FedFT) \cite{fedft} is a key enabler for achieving these objectives in the resource-constrained vehicular scenario. It leverages foundation model fine-tuning to reduce the number of training rounds, i.e., parameter transmissions, and employs Parameter-Efficient Fine-Tuning (PEFT) to significantly decrease the volume of transmitted parameters.

In pedestrian trajectory prediction literature, \cite{giuliari} and \cite{giuliari_journal} utilize Transformers, highlighting their superiority over Long Short-Term Memory (LSTM) models for capturing motion. To address independent movement modeling, \cite{Chen2024} introduces a spatial-temporal graph convolution network that captures complex social dynamics often neglected by baselines. In \cite{trajectory_prediction_federated}, trajectory prediction is solved via FL using a lightweight LSTM, offering enhanced privacy and performance over single-scene training. Furthermore, \cite{last_related_work} proposes a hierarchical, decentralized FL framework that achieves performance comparable to centralized FL while reducing communication overhead. Despite these advancements, these studies overlook the feasibility of such proposals in vehicular scenarios where radio resources are heavily constrained.

Our work focuses on the above setting and, unlike previous contributions, proposes to predict the pedestrian trajectory by leveraging the \textit{Chronos-2} foundation model \cite{ansari2025chronos2univariateuniversalforecasting}, trained with FedFT and the Low-Rank Adaptation (LoRA) technique. The clients that collaborate to train the deep neural network are Connected Vehicles (CVs), interacting with the orchestration server via Vehicle-to-Infrastructure (V2I) communications.  
In our proposal, the CVs are equipped with onboard sensors that collect environmental data and GPUs that extract the trajectories of the pedestrians encountered along the CV routes from these data.
Moreover, CVs train the model locally while transmitting LoRA updates to the server at a lower frequency than that of the training rounds.
We compare the performance of the approach against baseline methods, including centralized training, federated training, and full fine-tuning, and further investigate it for an increasing number of local training rounds per communication round.
Moreover, to quantify the cost of federated training when it is employed in a wireless setting and demonstrate its feasibility, we introduce an ad-hoc metric, which we call the  Radio Occupancy Time (RaOT).
The results demonstrate that our approach achieves high predictive accuracy while drastically reducing radio resource usage, making it an effective solution for real-world road safety applications in both 5G and 6G cellular systems.

The paper is organized as follows.
Section \ref{Problem Settings and Tools} introduces the scenario that sets the stage of our evaluations and the characteristics of the proposed FedFT scheme.
In Section \ref{Metrics}, the metrics that assess the accuracy of any AI model predicting the future movements of a pedestrian are recalled, along with the newly introduced metric that quantifies the communication cost.
The numerical results are provided in Section \ref{Numerical Results}. 
Section \ref{Conclusions} summarizes our findings.

\section{Problem Statement and Proposed Solution}
\label{Problem Settings and Tools}
In this work, we examine the problem of pedestrian trajectory prediction from the vehicular perspective and propose solving it through a federated approach among vehicles. 
The distinctive feature of the solution, which we term LoRA-FedFT, resides in the combined adoption of the Chronos-2 foundation model and LoRA to drastically reduce the communication cost, yet preserving the accuracy of the model in forecasting the future movements of the pedestrian.

\begin{figure}[h]
\includegraphics[trim=0.5cm 2cm 0cm 0cm, clip, width=0.95\linewidth]{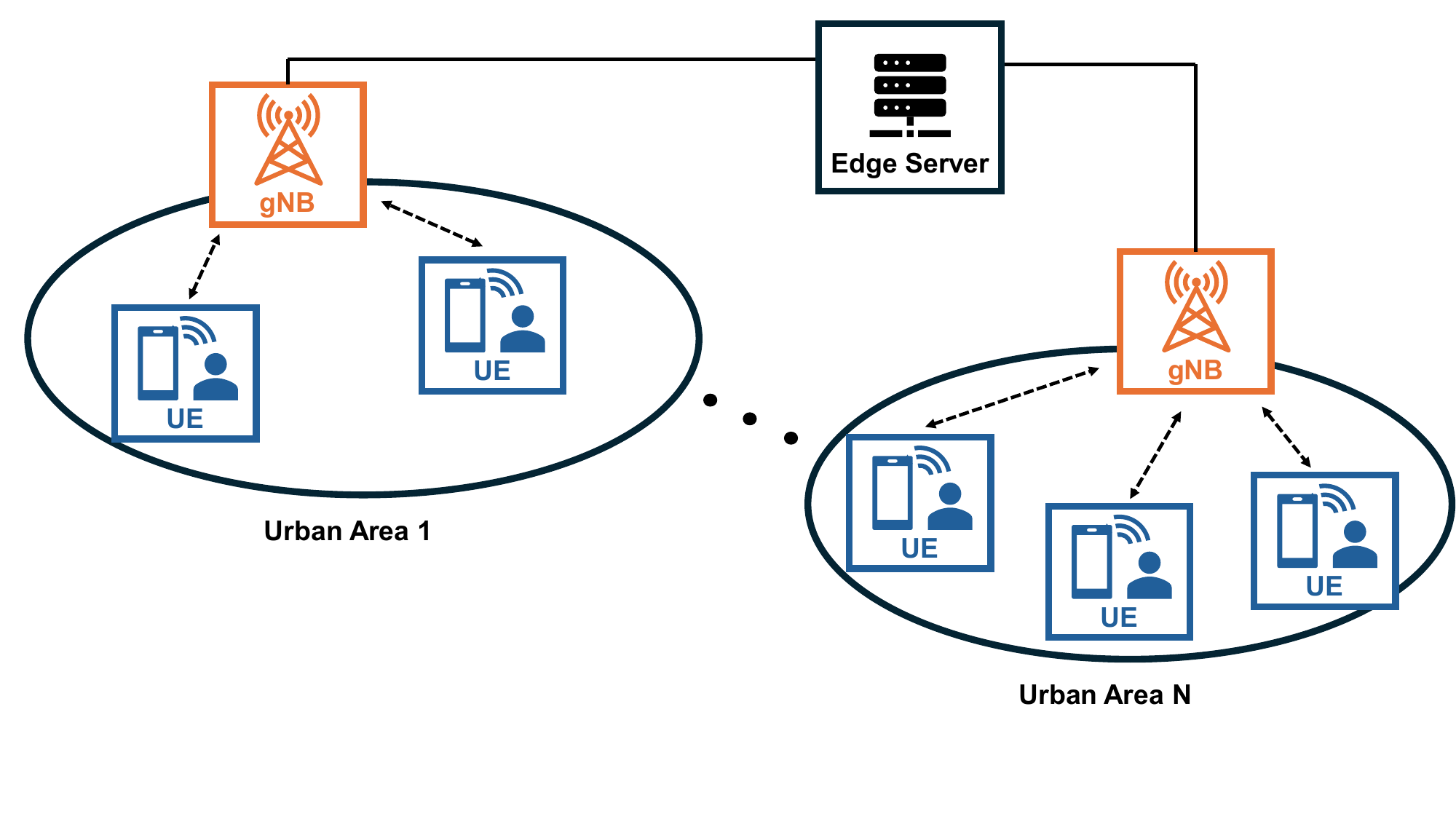}
\caption{The examined setting.}
\label{fig:scenario}
\end{figure}

\begin{figure}[htbp]
\noindent\hspace{-0cm}
\includegraphics[trim=6cm 2cm 2.5cm 2cm, clip, width=0.95\linewidth]{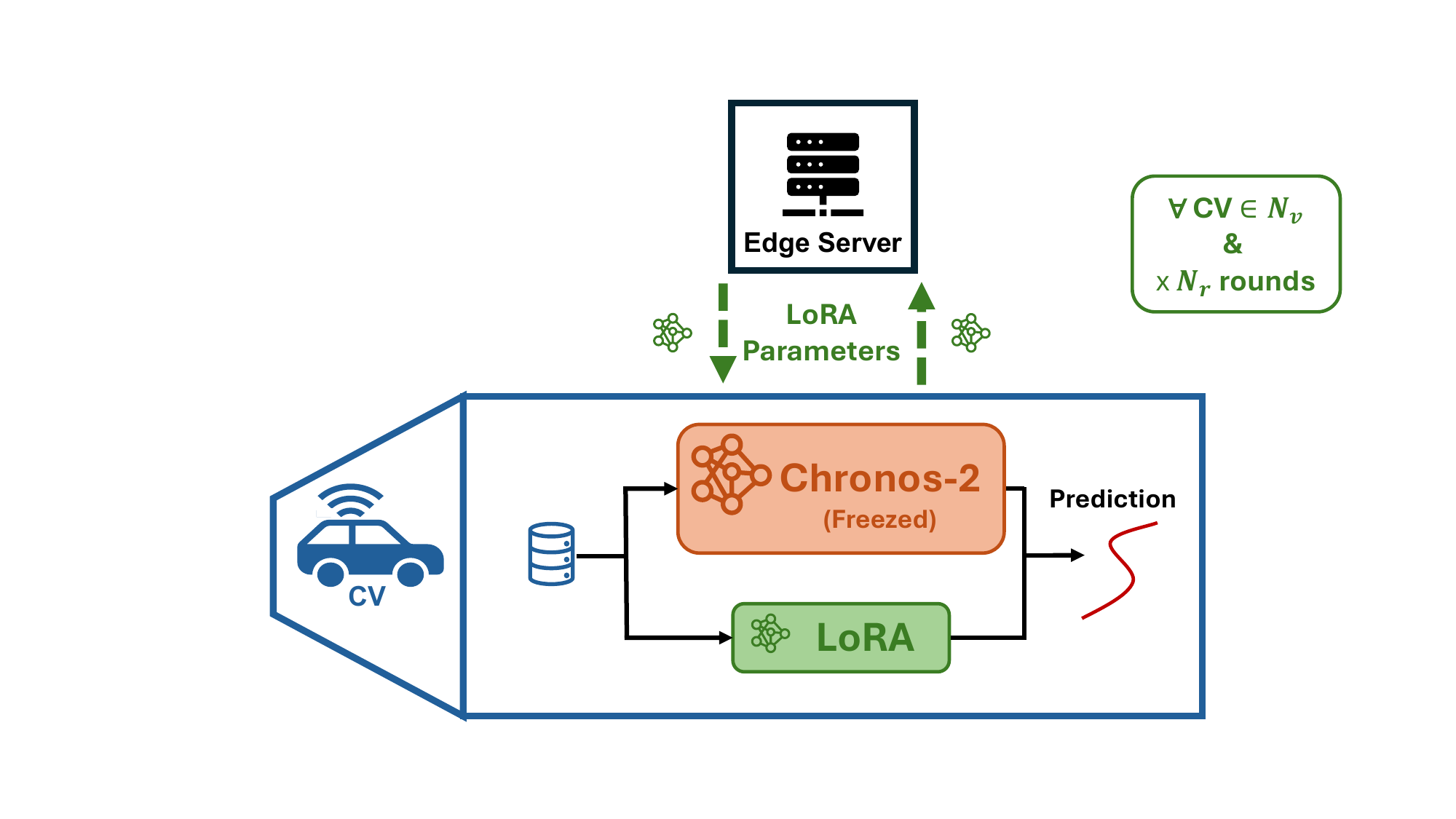}
\caption{Federated fine-tuning with LoRA.}
\label{fig:LoRA_architecture}
\end{figure}

In greater detail, we examine $N_v$ CVs traveling across different urban areas;
cellular coverage is provided to support the communication between each CV and the network edge, where the server orchestrating the update of the model's global parameters resides.
The system architecture is illustrated in Fig.\ref{fig:scenario}. 
We assume that the CVs collect information about pedestrians in their proximity via onboard sensors, e.g., cameras and lidars, and extract the pedestrian trajectories from the dataset they gradually build and store onboard. 
Every CV participates in the federated training and is therefore responsible for 
collaborating to fine-tune the foundation model using LoRA and transmitting the locally computed LoRA parameters over the wireless channel. 
An edge server of the cellular operator acts as the federated algorithm coordinator, 
and is therefore in charge of aggregating the LoRA parameters from the $N_v$ CVs, updating the model, and returning the CVs the
global model parameters, as the scheme in Fig.\ref{fig:LoRA_architecture} indicates.

The foundation model we choose is the state-of-the-art time-series forecasting model, i.e., the \textit{Chronos-2} foundation model released by Amazon in 2025 \cite{ansari2025chronos2univariateuniversalforecasting}. 
This model is an encoder-only transformer with an additional group attention module to leverage in-context learning, making it suitable for learning dependencies between the coordinates of the trajectories in the bidimensional space we consider. 
Its default version comprises more than $121$ million parameters, making it a model of modest size compared to Large Language Models. However, its size is problematic in the wireless federated scenario we examine: every CV should transmit hundreds of Megabytes per training round, which is unacceptable for the communication cost.
This is what motivates us to additionally introduce a Parameter Efficiency Fine Tuning strategy, namely, LoRA, which operates on small, low-rank matrices of rank $r$, while keeping the model parameters fixed. 
During the communication round, the CVs transmit only the LoRA parameters, whose size is typically some orders of magnitude lower than the model size, thereby substantially reducing the amount of data exchanged
over the radio channel. 

To further decrease the burden the scheme places on the channel, we introduce the possibility that the CVs transmit the local parameters to the aggregation server every $F$ training rounds, $F \geq 1$. Equivalently stated, we set $F$ to be equal to the ratio between $N_r$, the number of local training rounds, and $C_r$, the number of communication rounds, i.e., epochs, employed to wirelessly transfer the LoRA model updates to 
the network edge,
\begin{equation}
    \label{eq:F-definition}
    F= \frac{N_r}{C_r}
    \, , \quad F \geq 1
    \, .
\end{equation}

The aim is to determine the effects that different F values have on model accuracy and radio resource usage.

Lastly, observe that the output of \textit{Chronos-2} is multimodal, i.e., multiple trajectories are provided, each corresponding to a specific quantile level. It is worth mentioning that this model is usually trained considering the quantile regression objective:
\begin{equation}\label{eq:loss}
    \sum_{q\in \mathcal{Q}}\left( q\cdot \max\left( z-\hat{z}^q,0\right)+\left( 1-q\right) \cdot \max \left( \hat{z}^q-z,0\right) \right),
\end{equation}
where $\mathcal{Q}$ is the set of the quantile levels returned by \textit{Chronos-2}, $\hat{z}^q$ is the forecast at quantile level $q$, and $z$ is the corresponding target value. In this study, the loss function is assumed to coincide with the quantile regression objective.

\section{Task Metrics}\label{Metrics}
\subsection{Task-specific Metrics}
The effectiveness of a model in solving the task of trajectory prediction is assessed through the accuracy metrics reported below.

Let $J$ denote the number of trajectories in a set and $T$ the number of trajectory points to be predicted.
Then, the Average Displacement Error (ADE)
is defined as the
average $\ell_2$ distance between the predicted position $\hat{x}^j_t$ and the ground truth position $x^j_t$ over the entire prediction horizon and the set of trajectories, i.e., 
\begin{equation}
    ADE = \frac{1}{J\cdot T}\sum^{J}_{j=1}\sum^{T}_{t=1}\|\hat{x}^j_t-x^j_t\|_2
\end{equation}

The Final Displacement Error (FDE)
is the
average $\ell_2$ distance between the predicted final position $\hat{x}^j_T$ and the  ground truth $x^j_T$ over the set of trajectories, i.e.,
\begin{equation}
    FDE = \frac{1}{J}\sum^{J}_{j=1}\|\hat{x}^j_T-x^j_T\|_2
\end{equation}

Finally, the Miss Rate (MR)
at time
$\tau=n\cdot0.5s$, $n\in[1, T]$, is defined as the fraction of predictions whose $\ell_2$ distance from the ground truth exceeds a given threshold $\delta$, i.e., 

\begin{equation}
    MR = \frac{1}{J}\sum^{J}_{j=1}\mathbf{1}_{\{ \| \hat{x}^j_\tau-x^j_\tau\|_2 \ > \ \delta \}} \ .
    \label{eq:miss-rate}
\end{equation}

\begin{figure}[b]
\includegraphics[trim=4.3cm 3.5cm 5cm 4cm, clip, width=0.95\linewidth]{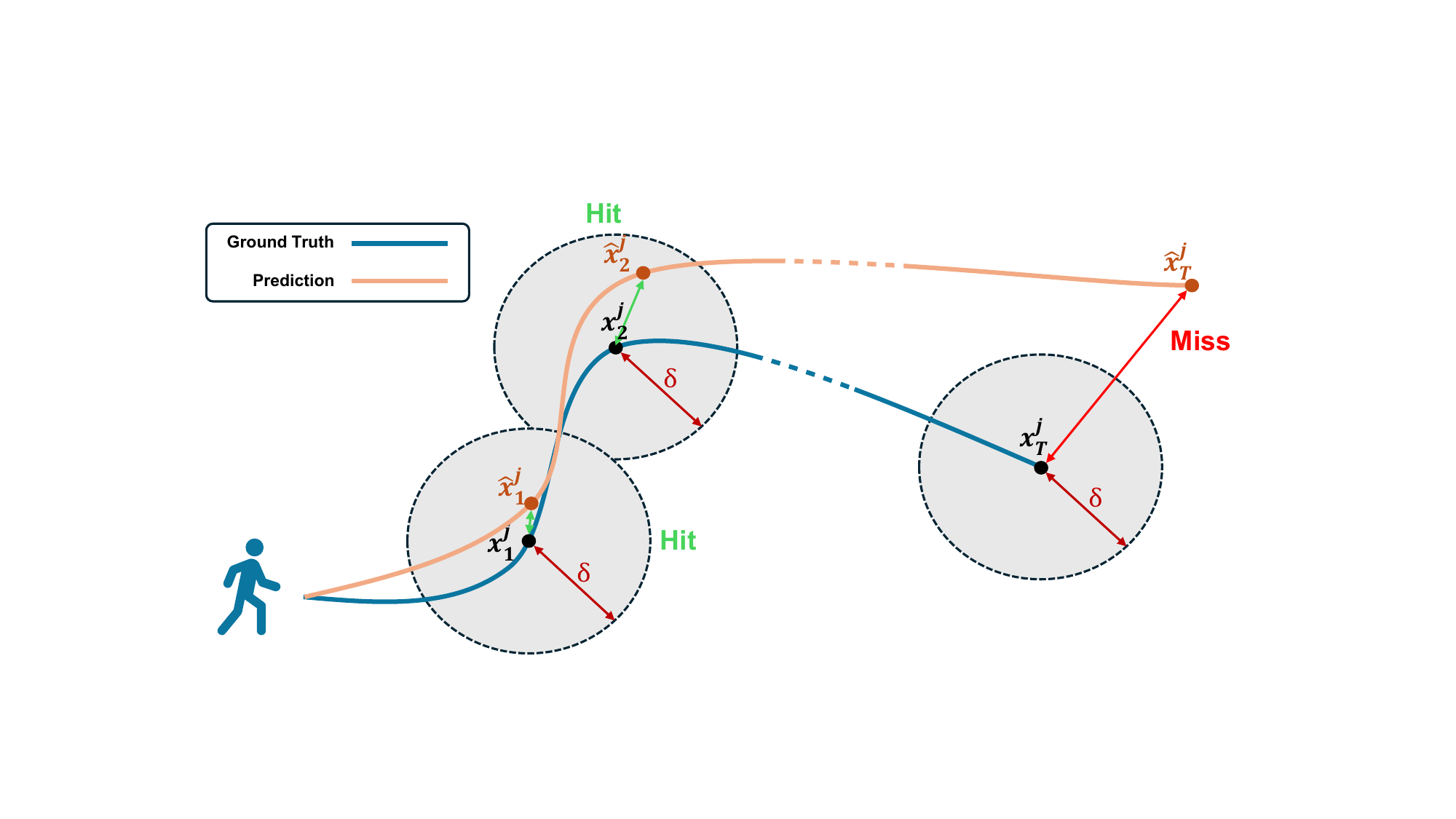}
\caption{Visual representation of the miss rate.}
\label{fig:miss_rate_visual}
\end{figure}

As visually depicted in Fig.\ref{fig:miss_rate_visual},  the $MR$ quantifies the probability that the coordinates $(\hat{x}_\tau, \hat{y}_\tau)$ of the estimated trajectory fall outside of a circle whose radius coincides with $\delta$.

\subsection{Communication Metric}
\label{Radio Occupancy Time and 6G impact}

To evaluate the communication cost of the proposed solution, we introduce a new metric, termed Radio Occupancy Time (RaOT), which we define as the total time during which the radio resources are occupied throughout the training procedure.
We neglect the contribution of downlink communications to RaOT, assuming that the base station, i.e., the 5G gNB, uses broadcast communications to return the model parameters to the CVs. Therefore, we focus exclusively on the uplink, where in each round, as many unicast communications as the number of CVs enable the transmission of the LoRA parameters from the CVs to their gNBs.  

Given $l_i$ denotes the constant size of the data the $i$-th CV transmits on the uplink per single communication round, and $D_i$ the supported data rate on the uplink, recalling that $C_r$ is the number of communication rounds each CV employs, then the RaOT is formally 
defined as:
\begin{equation}\label{eq:RaOT}
    RaOT = C_r \cdot \sum^{N_v}_{i=1}\frac{l_i}{D_i} 
    \, ,
\end{equation}
\noindent
which recalling \eqref{eq:F-definition} is equivalently written as
\begin{equation}\label{eq:RaOT2}
    RaOT = \frac{N_r}{F} \cdot \sum^{N_v}_{i=1}\frac{l_i}{D_i} 
    \, .
\end{equation}

In turn, $D_i$, the supported data rate on the uplink, depends on the cellular system
configuration. 
When a 5G architecture is examined, we estimate it as the maximum supported data rate, which is 
provided in 3GPP TS 38.306 \cite{datarate} as
\begin{equation}\label{eq:datarate}
    D_i=v_{layers}\cdot Q_m\cdot R\cdot \frac{N^{BW_i,\mu}_{PRB}\cdot 12}{T^\mu_S}\cdot \left( 1-OH\right)\text{,}
\end{equation}
where $v_{layers}$ denotes the number of MIMO layers, $Q_m$ the modulation order, and $R$ the code rate. 
Given that $\mu$ is the OFDM numerology, $N_{PRB}^{BW_i,\mu}$ denotes the number of Physical Resource Blocks (PRBs) within the available bandwidth $BW_i$ assigned to the $i$-th CV, while $T_s^\mu$ represents the OFDM symbol duration. Finally, $OH$ accounts for the overhead introduced by control signals.

To forecast how the RaOT requirement will change when a 6G system is considered, we refer to some of its expected characteristics, introduced in Release 20 and  
discussed in 6G scientific literature.
We focus on three distinct aspects of the 6G physical layer:
\subsubsection{MIMO Layers} 
In 3GPP TR 38.914 \cite{6gmimo}, it is indicated that the number of receive antennas in user equipment can reach up to 16. This suggests that, in 6G, vehicles may employ approximately twice as many antennas as in 5G.
\subsubsection{Modulation and Coding Scheme (MCS)} In the literature, many works elaborate on novel techniques to improve efficiency and robustness of the coding scheme as well as the adopted modulation. As notable examples, \cite{mcs_improvements2} introduces AI-based physical layer modules to enhance robustness and throughput, whereas in \cite{mcs_improvements1}, a joint design of channel coding and modulation is adopted for Probabilistically Shaped Polar-Coded Modulation (PS-PCM), enabling the selection of higher MCSs given the same Signal-to-Noise Ratio (SNR).
To reflect these possible enhancements in 6G, we make the hypothesis that for the same SNR, the selected 6G MCS has an index $I^{6G}_{MCS}$ equal to $I^{5G}_{MCS}+3$.
\subsubsection{Overhead} As indicated in \cite{Nokia6GFoundations}, the 6G Physical Downlink Control Channel (PDCCH) and Physical Uplink Control Channel (PUCCH) will be more flexibly and efficiently employed, mainly due to AI-based techniques. Therefore, we estimate the 6G overhead to be one-tenth of the 5G overhead.

\section{Numerical Results}\label{Numerical Results}

To emulate data collected by CVs, we considered 3 open-source datasets: NuScenes \cite{nuscenes}, UCY \cite{ucy}, and ETH \cite{eth}.
NuScenes comprises $1,000$ scenes collected by a vehicle-mounted sensor suite with six cameras, five radars, and one LiDAR. 
The data consists of annotated raw sensor measurements; they are heterogeneous in content, as various object types are detected along the road (e.g., passenger cars, vans, trucks, cyclists, pedestrians). From them, we extract only the pedestrian trajectories.
More accurately, NuScenes comprises $4$ datasets, each covering a different geographic region; we partition each dataset into $3$ equal-size subsets, resulting in $12$ subsets in total, each associated with a CV ideally roaming and collecting data within that area. 
The UCY dataset provides 6 smaller subsets, called Zara1, Zara2, Zara3, Students1, Students3, and Uni. 
ETH comprises 2 subsets, ETH and Hotel.
Both UCY and ETH datasets already consist of a collection of trajectories.
Because the size of the raw data from which the ETH and UCY trajectories were extracted is unavailable, we roughly estimate it by leveraging the known size of the Seaport subset and its number of trajectories.
Table \ref{tab:Datasets} summarizes the main characteristics of the $20$ available subsets, in terms of size and number of pedestrian trajectories that they include. 
\begin{table}[htbp]
\caption{Characteristics of the Datasets}
\label{tab:Datasets}
\begin{center}
\begin{tabular}{llcc} 
\toprule
\textbf{Dataset} & \textbf{Subset} & \textbf{N° Trajectories} & \textbf{Raw Size} $l_d$ [GB] \\
\midrule
\multirow{12}{*}{NuScenes} & Seaport 1        & 11'050 & 66 \\
                              & Seaport 2        & 11'049 & 66 \\
                              & Seaport 3        & 11'049 & 66 \\
                              & Holland Village 1 & 769    & 12 \\
                              & Holland Village 2 & 769    & 12 \\
                              & Holland Village 3 & 769    & 12 \\
                              & Queenstown 1     & 1'852  & 16.2 \\
                              & Queenstown 2     & 1'852  & 16.2 \\
                              & Queenstown 3     & 1'852  & 16.2 \\
                              & Onenorth 1       & 4'527  & 25.8 \\
                              & Onenorth 2       & 4'526  & 25.8 \\
                              & Onenorth 3       & 4'526  & 25.8 \\
\midrule
\multirow{6}{*}{UCY}       & Zara 1           & 1'976  & 11.8$^{\mathrm{a}}$ \\
                              & Zara 2           & 4'477  & 27$^{\mathrm{a}}$ \\
                              & Zara 3           & 1'760  & 10.5$^{\mathrm{a}}$ \\
                              & Students 1       & 11'691 & 70$^{\mathrm{a}}$ \\
                              & Students 3       & 8'988  & 54$^{\mathrm{a}}$ \\
                              & University       & 538    & 3.2$^{\mathrm{a}}$ \\
\midrule
\multirow{2}{*}{ETH}        & hotel            & 877    & 5.2$^{\mathrm{a}}$ \\
                              & eth              & 246    & 1.5$^{\mathrm{a}}$ \\
\toprule
\multicolumn{4}{l}{\small $^{\mathrm{a}}$Estimated size.}
\end{tabular}
\end{center}
\end{table}


The total number of collaborating CVs is $N_v=20$, and each is assigned one of the subsets described in Table \ref{tab:Datasets}. 
Through the LoRA-FedFT scheme,
the CVs collaboratively build a generalized model that takes as input the previous $8$ positions of the pedestrian, corresponding to the last $4$ seconds, and predicts the next $12$ positions over a future time horizon of $6$ seconds.


In the following figures, we focus on LoRA-FedFT with $F=1$, $F=2$, and $F=5$ to explore which value of $F$ is best. 
Moreover, we set the LoRA rank to $r = 8$ and built the matrices across all \textit{Chronos-2} linear layers.
To assess how effectively the proposed model learns from the datasets, we compare its performance with alternative solutions.
Namely, we consider the centralized training approach, where the edge server acts as the only computing node, receiving the raw subsets from the CVs;
the conventional federated training scheme, where the CVs extract the trajectories from their subsets and collaborate to train the \textit{Chronos-2} foundation model from scratch in a federated manner; 
finally, 
the federated full fine-tuning approach, where the CVs tune all model parameters starting from the \textit{Chronos-2} pre-trained set.
In all approaches we examined, we split each subset into $80\%$ for training and $20\%$ for validation.

 Table \ref{tab:training_rounds} reports $N_r$, the number of training rounds, and $C_r$, the number of communication rounds that the approaches execute, assuming all of them adopt early stopping with a patience of $\rho=10$, i.e., terminate when no further improvements in the ADE are observed for 10 consecutive communication rounds.
 Excluding the centralized approach, both the proposed LoRA-FedFT approach and the full fine-tuning scheme significantly reduce the number of communication rounds.

\begin{table}[htbp]
    \centering
    \caption{Number of Training and Communication Rounds}
    \begin{tabular}{lcc}
        \toprule
        \multirow{2}{*}{\textbf{Approach}} & \textbf{$N_r$} & \textbf{$C_r$} \\
            & $(\rho=10)$ & $(\rho=10)$ \\  
        \midrule
        LoRA-FedFT for  $F=1$ & 20 & 20 \\
        LoRA-FedFT for $F=2$ & 38 & 19 \\
        LoRA-FedFT for  $F=5$ & 75 & 15 \\
        Full fine-tuning & 16 & 16 \\ 
        Federated Training & 36 & 36 \\ 
        Centralized Training & 102 & 1 \\
        \bottomrule
    \end{tabular}
    \label{tab:training_rounds}
\end{table}

\begin{figure}[htbp]
    \noindent\hspace{-0.35cm}
    \includegraphics[trim=0cm 0cm 0cm 0cm, clip, width=1.0\linewidth]{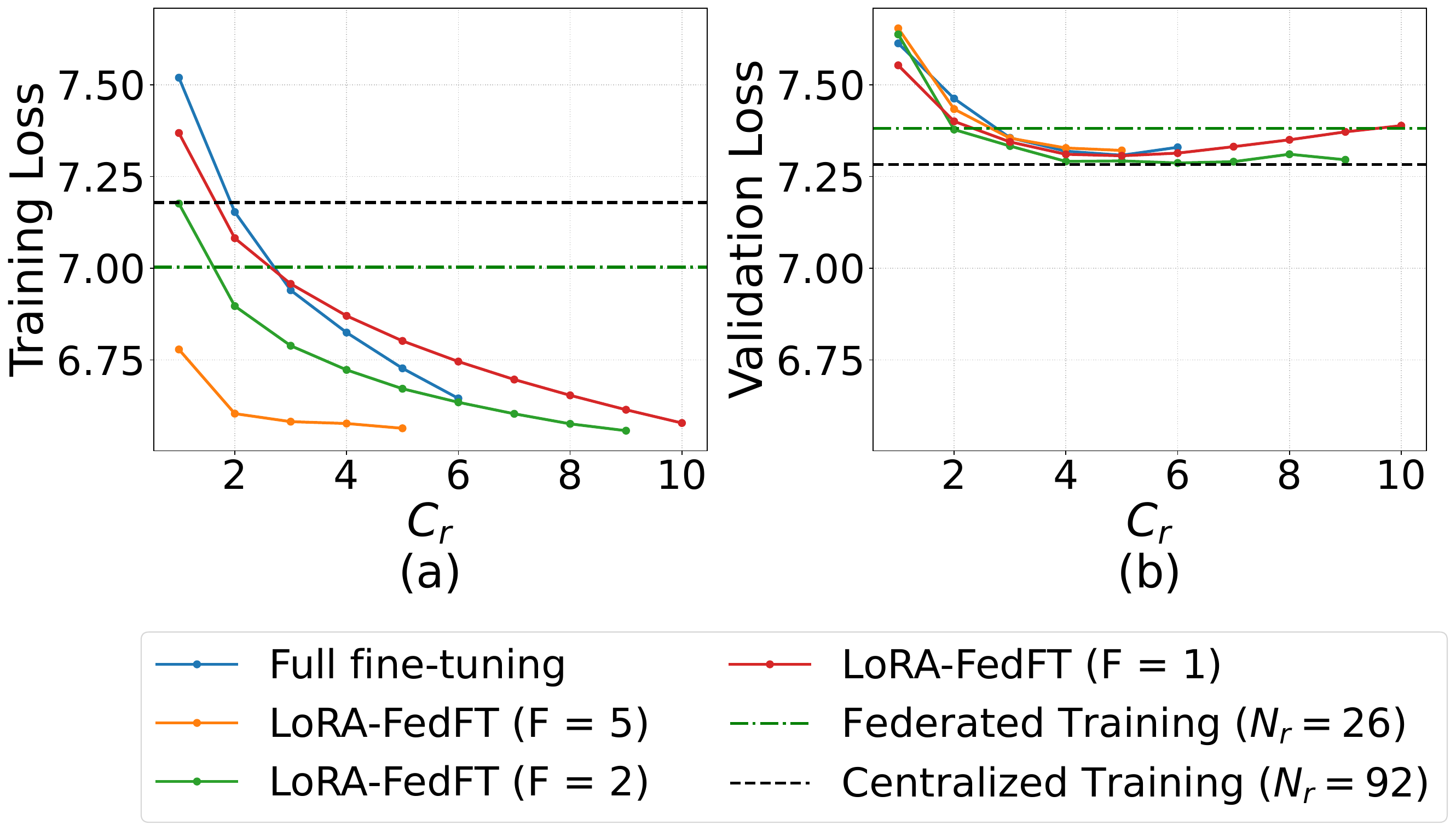}
    \caption{Training loss (a) and validation loss (b) during training.}
    \label{fig:loss}
\end{figure}

Fig.~\ref{fig:loss}(a) shows the training loss and Fig.~~\ref{fig:loss}(b) the validation loss as a function of $C_r$, determined by evaluating Eq. \eqref{eq:loss} over the training and validation data, respectively. The unit of measurement is dimensionless, as \textit{Chronos-2} calculates the losses on scaled inputs.
In this and the following figures, the curves of LoRA-FedFT with $F=1$, $F=2$, and $F=5$ are shown in orange, green, and red, respectively, and exclude the last $\rho$ communication rounds. The curve of the federated full fine-tuning approach is shown in blue.
For the centralized and federated training schemes, which act as baselines, 
we report the loss values at the round when the stopping mechanism detected the best ADE, at rounds $26$ and $92$, respectively, and insert two dashed horizontal lines in the figures.

\begin{figure}[htbp]
    \noindent\hspace{-0.45cm}
    \includegraphics[trim=0cm 0.cm 0cm 0cm, clip, width=1.0\linewidth]{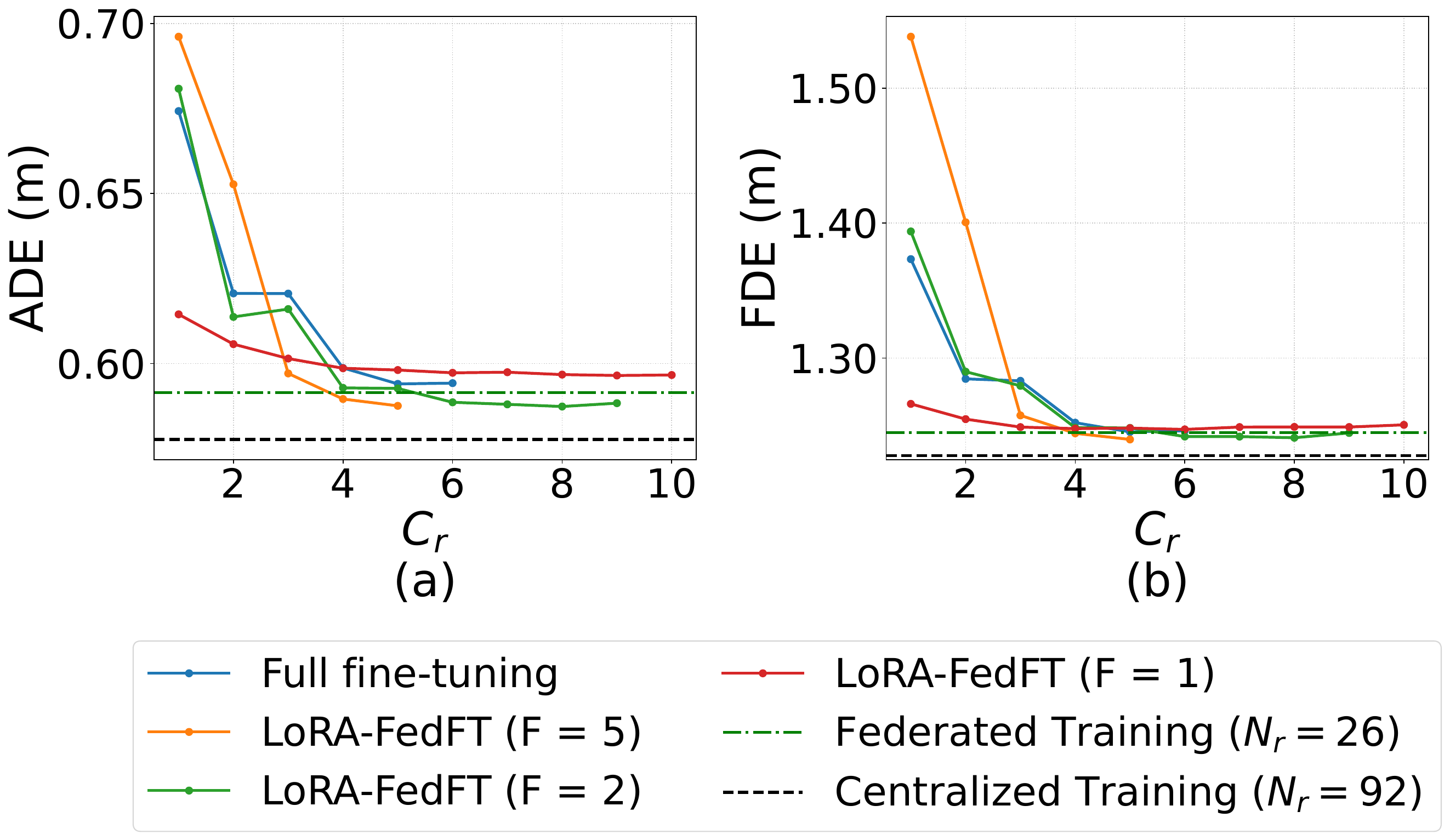}
    \caption{Displacement errors during training.}
    \label{fig:ade_fde}
\end{figure}

\begin{figure}[htbp]
    \noindent\hspace{-0.3cm}
    \includegraphics[width=1.0\linewidth]{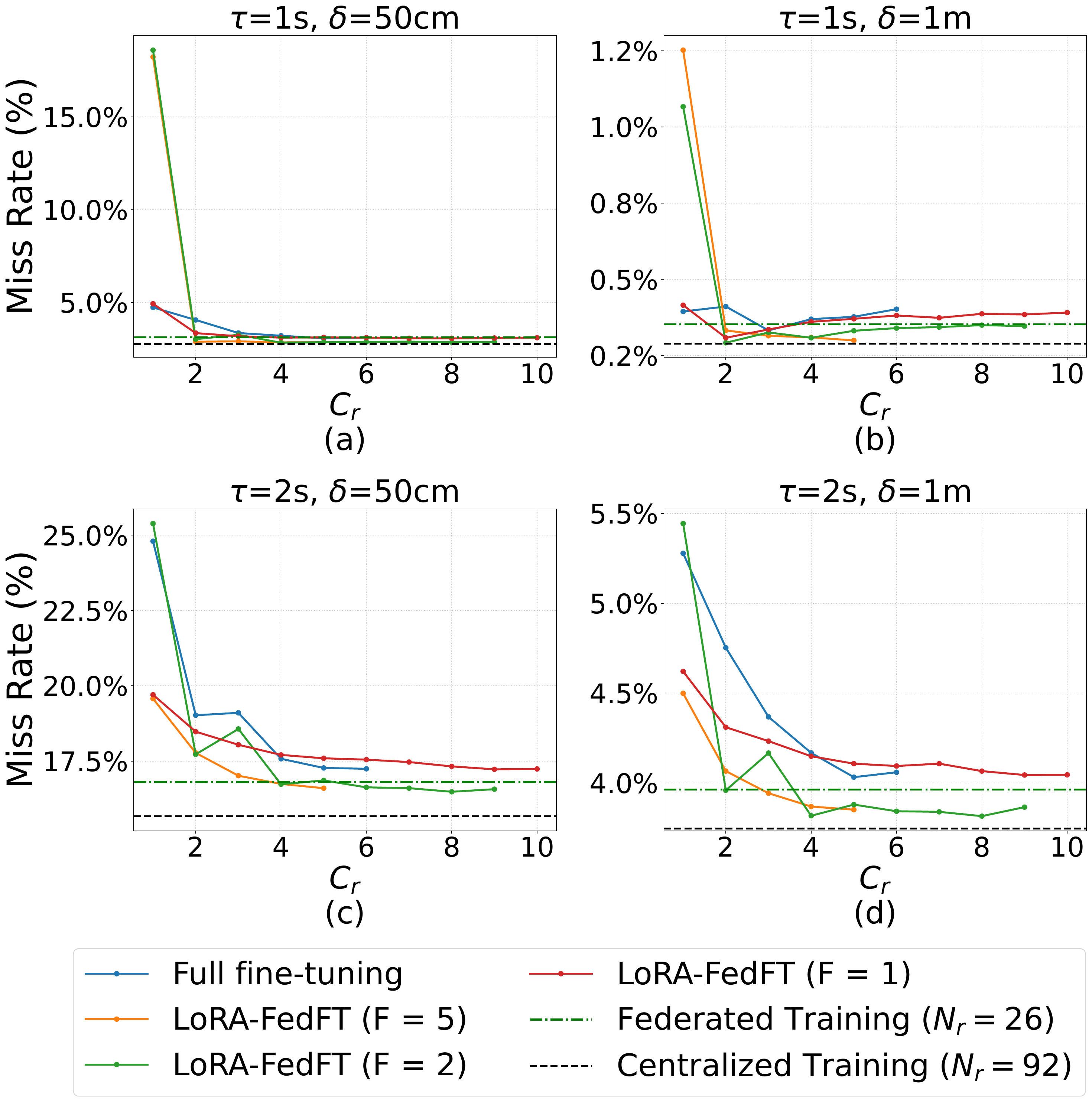}
    \caption{Miss rates during training.}
    \label{fig:miss_rates}
\end{figure}

All federated solutions show a lower training loss than the centralized solution when $C_r>2$. Each CV learns easily from its local dataset, whose trajectories share similar characteristics, yielding a clear optimization path. This phenomenon becomes more pronounced as the parameter $F$ increases, i.e., when each client refines the model on its local dataset for more rounds before server averaging.
In the centralized setting, this is harder because very different trajectories are used to find the optimal path down the loss function. 
In terms of validation loss, fine-tuning approaches reach values similar to the centralized solution, with a slight overfitting except for LoRA-FedFT with $F=5$.
Next, Figs.\ref{fig:ade_fde}(a) and (b) show the ADE and FDE evolution, respectively. Errors are portrayed as a function of the communication rounds $C_r$ and evaluated employing the $0.5$ quantile, i.e., the median.  
Consistent with the previous observations, centralized training achieves the lowest error values, whereas the fine-tuning approaches (full and LoRA-based) closely follow the federated training errors, with the schemes $F=2$ and $F=5$ showing slightly better values, and the latter converging more rapidly. 
In greater detail, the ADE of LoRA-FedFT for $F=2$ and $F=5$ are slightly below $0.59$ meters, i.e., they are only $1$~cm higher than the average error guaranteed by the centralized training, and the same remark applies to their FDE.
Next, Table \ref{tab:ade/fde} reports the ADE and FDE achieved by the LoRA-FedFT models on every single subset. The approach using $F=2$ and $F=2$ show slightly better performance across nearly all subsets. Combining Table \ref{tab:ade/fde} and Fig. \ref{fig:ade_fde}, the method that employs $F=5$ achieves the best accuracy among the federated approaches, matching $F=2$ while saving 4 communication rounds. The scheme with $F=1$ falls short in both performance and communication efficiency since it converges more slowly.
To determine the miss rate, we again considered the $0.5$ quantile trajectory. Figs.\ref{fig:miss_rates} (a) and (b) show the miss rates at time $\tau=2$~s when the threshold in \eqref{eq:miss-rate} is $\delta=50$~cm and $1$~m, respectively.
Figs.\ref{fig:miss_rates} (c) and (d) report the miss rate at $\tau=1$~s for the same $\delta$  values as before. 
The same trends identified for the ADE and FDE are observed, with LoRA-FedFT adopting $F=2$ and $F=5$ performing better than the other federated solutions and approaching in some cases the centralized one. 
In particular, in Fig.~\ref{fig:miss_rates}(b),
LoRA-FedFT nearly matches centralized training performance, with a miss rate of $0.3\%$ for either $F=2$ or $F=5$, demonstrating that the proposed approach effectively estimates the future evolution of a pedestrian trajectory.

\subsection{Communication Cost}
To evaluate and compare the communication cost of the alternative solutions, we assume that in centralized training, the CVs do not possess sufficient processing capability to extract the trajectories from the data collected through the onboard sensors; hence, they have to deliver their raw data to the edge server, whose sizes are reported in Table \ref{tab:Datasets}. 
For both federated training and full fine-tuning, the data to be transmitted in each communication round comprises more than 121M model parameters, whereas LoRA-FedFT transmits only 2M parameters, representing the LoRA matrices. In this work, each parameter is represented as \texttt{float32}, i.e., 4 bytes, resulting in $l_i =477.91$MB and $l_i=7.8$MB for the full model parameters and LoRA parameters, respectively.
\begin{table}[htbp]
    \centering
    \setlength{\tabcolsep}{8pt}
    \caption{Results: ADE / FDE (meters)}
    \begin{tabular}{lccc}
        \toprule
        \multirow{2}{*}{\textbf{Subset}} & \textbf{LoRA-FedFT} & \textbf{LoRA-FedFT} & \textbf{LoRA-FedFT} \\
           & \textbf{(F=1)} & \textbf{(F=2)} & \textbf{(F=5)} \\
        \midrule
        Seaport 1 & 0.69 / 1.41 & 0.68 / 1.40 & \textbf{0.67} / \textbf{1.38} \\
        Seaport 2 & 0.71 / 1.48 & \textbf{0.70} / \textbf{1.47} & \textbf{0.70} / \textbf{1.47} \\
        Seaport 3 & 0.64 / 1.32 & \textbf{0.63} / 1.32 & \textbf{0.63} / \textbf{1.31} \\
        Holland V. 1 & 0.68 / 1.29 & 0.67 / 1.30 & \textbf{0.66} / \textbf{1.27} \\
        Holland V. 2 & 0.98 / 2.15 & 0.96 / 2.12 & \textbf{0.95} / \textbf{2.09} \\
        Holland V. 3 & 0.72 / \textbf{1.50} & \textbf{0.71} / \textbf{1.50} & 0.72 / \textbf{1.50} \\
        Queenst. 1 & 0.53 / 1.08 & \textbf{0.52} / \textbf{1.06} & 0.53 / 1.07 \\
        Queenst. 2 & 0.75 / \textbf{1.56} & \textbf{0.74} / \textbf{1.56} & \textbf{0.74} / 1.57 \\
        Queenst. 3 & 0.59 / 1.24 & \textbf{0.57} / \textbf{1.23} & 0.58 / 1.24 \\
        Onenorth 1 & 0.66 / 1.37 & \textbf{0.65} / 1.37 & \textbf{0.65} / \textbf{1.36} \\
        Onenorth 2 & 0.65 / 1.33 & 0.64 / 1.32 & \textbf{0.63} / \textbf{1.30} \\
        Onenorth 3 & 0.60 / 1.23 & \textbf{0.59} / 1.22 & \textbf{0.59} / \textbf{1.21} \\
        Zara 1 & 0.45 / 0.96 & \textbf{0.44} / \textbf{0.95} & \textbf{0.44} / 0.97 \\
        Zara 2 & 0.36 / \textbf{0.77} & \textbf{0.35} / \textbf{0.77} & \textbf{0.35} / \textbf{0.77} \\
        Zara 3 & 0.49 / 1.07 & \textbf{0.47} / \textbf{1.05} & 0.48 / 1.06 \\
        Students 1 & \textbf{0.48} / \textbf{1.03} & \textbf{0.48} / \textbf{1.03} & \textbf{0.48} / \textbf{1.03} \\
        Students 3 & 0.63 / 1.37 & \textbf{0.62} / \textbf{1.36} & \textbf{0.62} / \textbf{1.36} \\
        University & 0.54 / 1.20 & \textbf{0.53} / 1.17 & \textbf{0.53} / \textbf{1.16} \\
        hotel & 0.30 / 0.55 & \textbf{0.28} / \textbf{0.52} & \textbf{0.28} / \textbf{0.52} \\
        eth & \textbf{0.91} / \textbf{1.77} & 0.94 / 1.83 & \textbf{0.91} / 1.78 \\
        \midrule
        \textbf{Weigh. Avg.} & 0.60 / 1.25 & \textbf{0.59} / \textbf{1.24} & \textbf{0.59} / \textbf{1.24} \\
        \bottomrule
    \end{tabular}
    \label{tab:ade/fde}
\end{table}
\begin{figure}[b]
    \noindent\hspace{0cm}
    \includegraphics[trim=0cm 0.5cm 0.5cm 0cm, clip, width=1.0\linewidth]{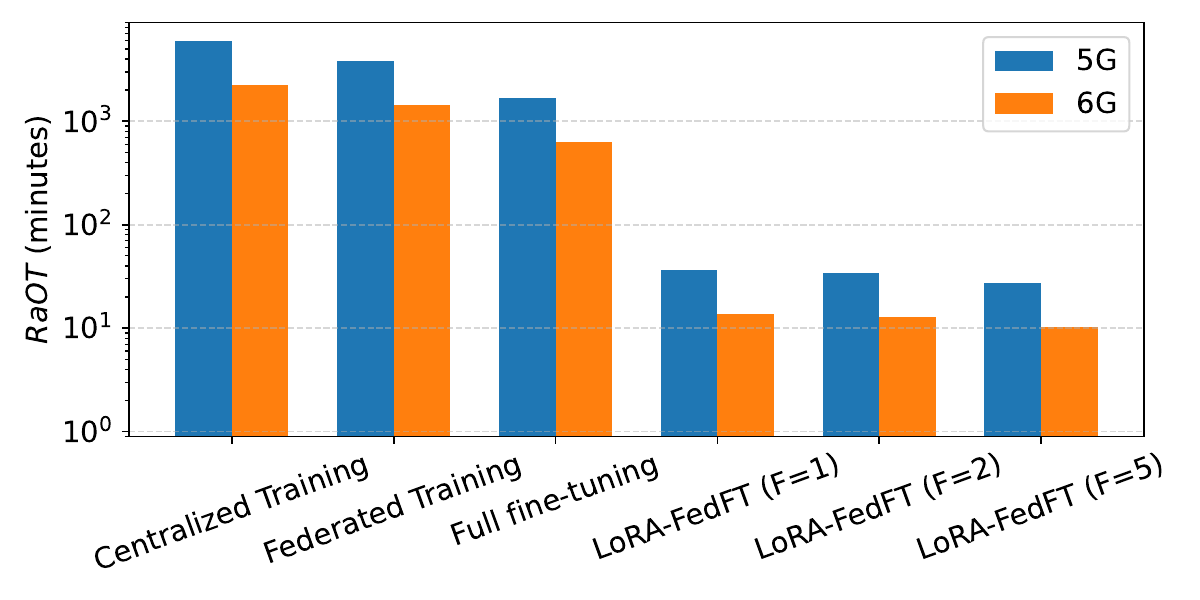}
    \caption{Radio communication costs.}
    \label{fig:comm_cost}
\end{figure}
We assume the entire 5G infrastructure uses a $BW=20$~MHz bandwidth to support uplink communications for all 20 CVs, i.e., $BW_i=1$~MHz per CV. 
Furthermore, as an illustrative example, the OFDM numerology is set to $\mu=0$, which results in a total number of $N_{PRB}^{BW,\mu}=5$ PRBs per CV and a symbol duration $T_s^\mu=71.4$~$\mu$s. Additionally, we set the MCS index to $I_{MCS}^{5G}=22$, corresponding to a modulation order $Q_m=6$ and a code rate $R=0.702$.
We consider CVs to be high-end 5G mobile devices, capable of establishing $v_{layers}=4$ MIMO layers.
Finally, we set $OH=0.08$.
We calculate the corresponding RaOT values considering the communication rounds $C_r$ in Table \ref{tab:training_rounds}, and report them in Fig.\ref{fig:comm_cost}: the blue rectangles corresponding to 5G highlight that both the centralized and federated training require radio resources for a cumulative time approximately equal to $4$ days $4$ hours, and $2$ days $15$ hours, respectively.
Full fine-tuning reduces the RaOT to  $1$ day and $4$ hours. Yet, the system remains resource-hungry.
With the LoRA-FedFT approach, the RaOT drops to $27$ minutes for $F=5$, showing the solution is fully feasible in 5G.
When examining 6G, we assume $R=0.853$, $ v _ {layers} = 8$, and an overhead value $OH=0.008$.
Accordingly, we obtain the RaOT values displayed in orange.
They pleasingly reduce to $12$ minutes and $54$ seconds for LoRA-FedFT $F=2$, and to $10$ minutes and $11$ seconds for $F=5$.
Without aiming to identify the actual RaOT decrease expected in 6G, these values demonstrate that the proposed scheme is a strong candidate for solving complex tasks in the vehicular setting, enabling the federated implementation of foundation models in current and future cellular networks.

\vspace{-0.2truecm}
\section{Conclusions}\label{Conclusions}
\vspace{-0.2truecm}
In this paper, we demonstrated that the LoRA-FedFT approach can successfully solve the problem of pedestrian trajectory prediction in the vehicular environment,
remarkably reducing the number of V2I communications, and established the feasibility of the proposal both in 5G and 6G networks.
We also showed that setting the update frequency of the model's global parameters to one-fifth of the local training-round frequency further reduces the communication cost of the proposed federated scheme while preserving prediction accuracy.

\vspace{-0.48truecm}
\bibliographystyle{ieeetr} 
\bibliography{bib/references_ccnc}

\end{document}